\documentclass[11pt,twoside]{article}

\usepackage{asp2006}
\usepackage{epsf}

\begin{document}
\title{FIRST `Winged' and `X'-shaped Radio Source Candidates} 
\author{C. C. Cheung\altaffilmark{1}, A. Springmann\altaffilmark{2}}   

\affil{Kavli Institute for Particle Astrophysics and Cosmology\\ 
Stanford University, Stanford, CA 94305, USA}

\altaffiltext{1}{Jansky Postdoctoral Fellow of the National Radio
Astronomy Observatory. The NRAO is operated by Associated Universities, 
Inc. under a cooperative agreement with the NSF.} 
\altaffiltext{2}{Astronomy Department, Wellesley College, Wellesley, MA 
02481. Supported by the U.S. Department of Energy through the SULI program 
at the Stanford Linear Accelerator Center.}

\begin{abstract} A small number of double-lobed radio galaxies are found
with an additional pair of extended low surface brightness
`wings' of emission giving them a distinctive `X'-shaped appearance.  One 
popular explanation for the unusual morphologies posits that the central
supermassive black hole (SMBH)/accretion disk system underwent a recent 
realignment; in a merger scenario, the active lobes mark the
post-merger axis of the resultant system (e.g., Merritt \& Ekers 2002).
However, this and other interpretations are not well tested on the few
(about one dozen) known examples. In part to remedy this deficiency,
a large sample of winged and X-shaped radio sources is being compiled for
a systematic study.  An initial sample of 100 new candidates is described
as well as some of the follow-up work being pursued to test the different 
scenarios. \end{abstract}

\vspace{-0.2in}
\section{FIRST Identifications and Work in Progress}  

To compile the sample, image fields from the VLA-FIRST survey (Becker et
al. 1995) containing components bright and extended enough to judge the
source morphologies were inspected by-eye. This gave an initial 100
candidates with extended winged emission (Fig.~1; Cheung 2006).  Compared
to previously known examples (e.g., Lal \& Rao 2006), the new candidates
are systematically fainter ($\sim$10$\times$) and more distant
($z$$\lower.5ex\hbox{$\; \buildrel > \over \sim \;$}$0.3).  New optical
spectroscopic observations are identifying many of the fainter, more
distant optical hosts. 

Most candidates have clear winged emission and higher resolution VLA
observations of initially $\sim$40 sources have been obtained to confirm
the morphological identifications. Of the candidates, enough are legitimate
X-shaped sources (conventionally, those with wing to lobe extents of
$>$0.8:1) to more than double the number known. Lower frequency GMRT
observations of selected objects are being pursued to map any spectral
structure to estimate the particle ages in the wings to test formation
scenarios (e.g., Dennett-Thorpe et al. 2002). 

We examined the host galaxies of about a dozen new and previously known
examples with available SDSS images (54 sec exposures) to quantify any
asymmetry in the surrounding medium as required by hydrodynamic wing
formation models (e.g., Capetti et al. 2002). Most of the galaxies are
highly elliptical with the minor axes roughly aligned with the wings,
consistent with the findings of Capetti et al. for a similarly sized
sample.  However, we found smaller ellipticities ($\epsilon$$<$0.1) in at
least two examples, 3C192 and B2~0828+32, confirming previous studies of
these hosts (Smith \& Heckman 1989; Ulrich \& R{\" o}nnback 1996). 
``Round" hosts are not necessarily incompatible with the hydrodynamic
picture as observed $\epsilon$ values can be lowered by projection.
This should be investigated more thoroughly with a dedicated
host galaxy imaging program.

\begin{figure*}[t]
\plotone{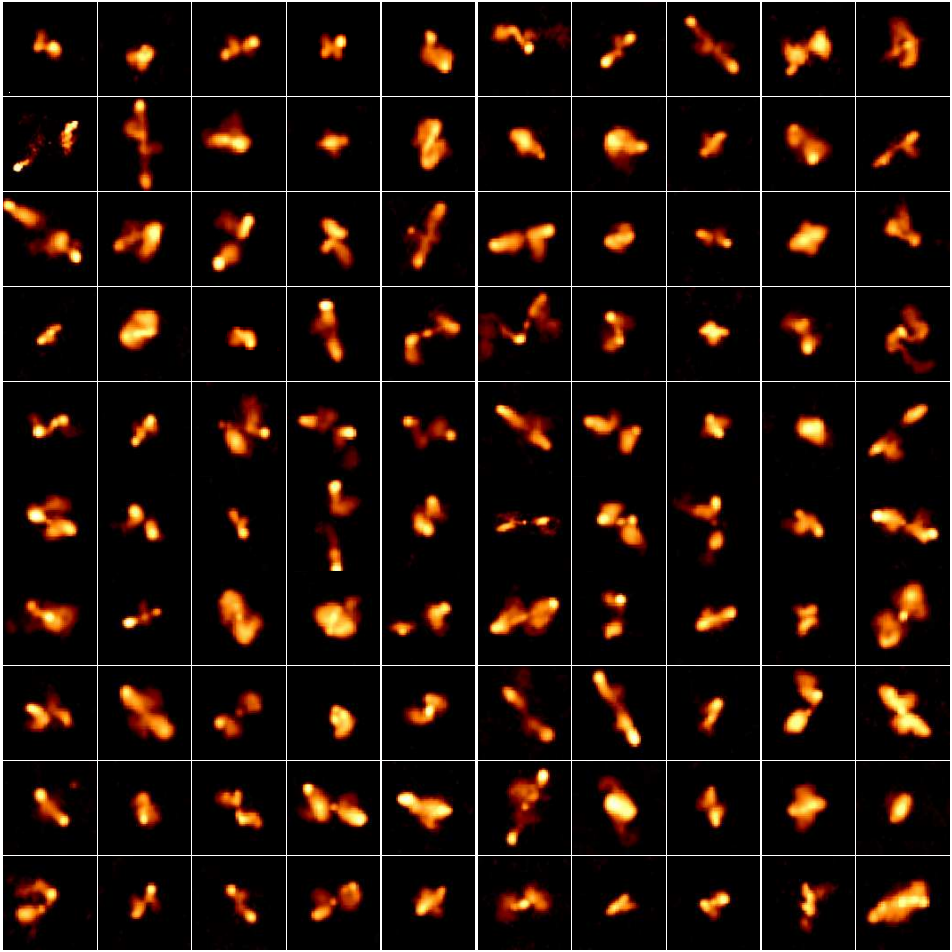}
\vspace{-0.075in}
\caption{VLA-FIRST $\lambda$20cm images 
($\sim$100\hbox{$^{\prime\prime}$}$\times$100\hbox{$^{\prime\prime}$}) of 
the 100 new winged and 
X-shaped radio source candidates at 5.4\hbox{$^{\prime\prime}$}\ resolution 
from Cheung (2006).} \end{figure*}

%Thanks are due to our collaborators on the follow-up work (H. Landt, S. 
%Healey, G. Kleijn, A. Jordan, and D. Lal).


\vspace{-0.075in} 

\begin{thebibliography}{}
\vspace{-0.075in}
\bibitem[Becker et al.(1995)]{bec95} Becker, R.H., White, R.L., \& Helfand, D.J.\ 1995, ApJ, 450, 559
%\bibitem[Begelman et al.(1980)]{beg80} Begelman, M.C., Blandford, R.D., \& Rees, M.J.\ 1980, Nature, 287, 307
\bibitem[Capetti et al.(2002)]{cap02} Capetti, A., et al.\ 2002, A\&A, 394, 39
\bibitem[Cheung(2006)]{che06} Cheung, C.C.\ 2006, AJ, submitted
\bibitem[Dennett-Thorpe et al.(2002)]{den2} Dennett-Thorpe, J. et al. 2002, MNRAS, 330, 609
\bibitem[Lal \& Rao(2006)]{lal06} Lal, D.V., \& Rao, A.P.\ 2006, MNRAS, in press (astro-ph/0610678)
\bibitem[Merritt \& Ekers(2002)]{mer02} Merritt, D., \& Ekers, R.D. 2002, Science, 297, 1310
\bibitem[Smith \& Heckman(1989)]{smi89} Smith, E.P., \& Heckman, T.M. 1989, ApJS, 69, 365
\bibitem[Ulrich \& R{\" o}nnback(1996)]{ulr96} Ulrich, M.-H., \& R{\" o}nnback, J. 1996, A\&A, 313, 750 

\end{thebibliography}
\end{document}